 \documentclass[aps,prb,twocolumn,superscriptaddress,showpacs]{revtex4}

\usepackage{latexsym} 
\usepackage[dvips]{graphicx}
\usepackage[applemac]{inputenc} 
\usepackage{amsmath}
\usepackage{tabularx} 
\usepackage{tipa}

\begin{document}

\title{Intrinsic avalanches and collective phenomena in a Mn(II)-free radical ferrimagnetic chain}

\author{E. Lhotel}
\affiliation{Institut N\'eel, Dpt MCBT, CNRS \& Universit\'e Joseph Fourier, BP166, F-38042 Grenoble Cedex 9, France} 
\email[Corresponding author: ]{elsa.lhotel@grenoble.cnrs.fr}
\author{D. B. Amabilino}
\author{C. Sporer}
\affiliation{Institut de Ci\`encia de Materials de Barcelona (CSIC), Campus Universitari, 08193 Bellaterra, Spain}
\author{D. Luneau}
\affiliation{Universit\'e Claude Bernard-Lyon 1, B\^at 305, 69622 Villeurbanne Cedex, France}
\author{J. Veciana}
\affiliation{Institut de Ci\`encia de Materials de Barcelona (CSIC), Campus Universitari, 08193 Bellaterra, Spain}
\author{C. Paulsen}
\affiliation{Institut N\'eel, Dpt MCBT, CNRS \& Universit\'e Joseph Fourier, BP166, F-38042 Grenoble Cedex 9, France}

\begin{abstract}
Magnetic hysteresis loops below 300 mK on single crystals of the Mn(II) - nitronyl nitroxide free radical chain (Mn(hfac)$_2$({\it R})-3MLNN) present abrupt reversals of the magnetization, or avalanches. We show that, below 200 mK, the avalanches occur at a constant field, independent of the sample and so propose that this avalanche field is an intrinsic property. We compare this field to the energy barrier existing in the sample and conclude that the avalanches are provoked by multiple nucleation of domain-walls along the chains. The different avalanche field observed in the zero field cooled magnetization curves suggests that the avalanche mechanisms are related to the competition between ferromagnetic and antiferromagnetic orders in this compound. 

\end{abstract}

\pacs{75.50.Xx, 75.10.Pq, 75.60.-d, 75.60.Jk}	

\maketitle

\section{Introduction}
Magnetic avalanches, that is to say abrupt magnetization reversals, often appear in low temperature hysteresis loops  \cite{Prejean80_2, Paulsen95, Uehara86}. In some cases, they are just giant Barkhausen jumps and so give information about defaults in the system.  An interesting situation arises when avalanches are intrinsic to the system. Then they can be a powerful tool for understanding the underlying interactions, and the mechanisms of nucleation and domain-wall propagation. 

Much interest has been devoted to the study of spin dynamics in molecular nano-cluster  zero-dimensional (0D) systems and one-dimensional (1D) or quasi-1D systems. In the latter, because of the strong interactions within the chains, the magnetic relaxation is expected to present an unusual behavior.  In particular, slow dynamics have been investigated in single chain magnets  \cite{Caneschi01, Clerac02} as well as in ordered spin chains \cite{Ovchinnikov06}. Recently, unexpected resonant effects, attributed to quantum tunneling, have been observed in the three-dimensional (3D) ordered spin chain [(CH$_3$)$_3$NH]CoCl$_3\cdot$2H$_2$O (CoTAC) \cite{Lhotel06b}. 

The study of avalanches in such systems can provide additional information about the reversal processes.  A  pertinent question concerns the starting flip of a  population of spins at the origin of the avalanche. A quantum origin was proven in the early results on the Mn$_{12}$ac single molecule magnet (SMM) \cite{Sessoli93} which appears as an emblematic example of intrinsic avalanches in molecular compounds. The avalanches were shown to occur around the resonance field and were thus attributed to resonant quantum tunneling of the magnetization which increases the spin flipping rate \cite{Paulsen95}.   

Another important question concerns the propagation of the avalanche and thus the associated time scale required for a macroscopic magnetization reversal to develop. For example, recent avalanches studies in  Mn$_{12}$ac have shown that the magnetization reversal is not uniform inside the sample \cite{Bal04} and that the avalanches propagate at a constant velocity, requiring a threshold energy \cite{Suzuki05_1}.
In most cases, the propagation of the avalanche can be attributed to thermally assisted phenomena. At low temperature, it is often difficult for the heat released by the spin flipping  process to be dissipated in the sample and to be absorbed by the external thermal bath (due to poor thermal coupling of the sample to the environment and/or low thermal conductivity). In this scenario, the local heating due to the flipping of a small group or  cluster of spins is sufficient to heat the neighboring spins to a temperature where thermal activation is efficient, thus enabling them to flip, which in turn will heat their neighbors, and so on.

To know whether or not the latter mechanism governs the avalanche, two experimental tests can be done. One is simply to vary the sample size and the coupling to the heat bath. Indeed, the use of quite small samples allows better thermal homogeneity and avoids excessive overheating during a magnetization reversal process, thus suppressing the avalanche. For example, when a small crystal of Mn$_{12}$ac was directly immersed in  liquid $^4$He-$^3$He inside the mixing chamber of the dilution refrigerator, the sudden avalanches were replaced by anomalies in the hysteresis loop (called steps), which are the signature of the relaxation of the magnetization by quantum tunneling  \cite{Perenboom98}. Another test is to vary the field sweeping rate. During a field ramp, a certain amount of small stochastic spin flips may occur, which are due to defaults or impurities in the sample and may be considered extrinsic. However, if enough are present and if the field is ramped too fast, then the combined heat released from all of these processes during a short time can trigger a thermal avalanche. The field at which this kind of avalanche occurs will depend on the temperature and ramping rate in a very complicated way. On the other hand, by ramping more slowly, the heat will have time to dissipate and the avalanches will disappear. This was the case at low temperature for  CoTAC \cite{Lhotel06b}, where, for example, avalanches occurred at field values as low as 100 Oe  when the field was ramped at 75 Oe.s$^{-1}$.  By simply ramping at a slower rate of approximately 1 Oe.s$^{-1}$, avalanches were suppressed, and we were able to measure relaxation in fields up to 2000 Oe and, in particular, explore the resonant tunneling near 1000 Oe.

In the present paper, we are interested in the avalanche phenomena in a metallo-organic system, a Mn (II)-free radical ferrimagnetic chain. Contrary to the above  systems, here, we show that the avalanches are intrinsic both in their origin and their propagation. Indeed, the avalanches are found to be sample independent at low temperature and to occur at a well-defined internal field. They do not disappear even by using very slow field sweeping rates, down to $0.04$ Oe.s$^{-1}$, and with samples as small as 10$^{-4}$ mm$^3$. 

In Sec. \ref{exp}, we briefly describe the samples, review some pertinent experimental features, and summarize the structural and magnetic properties from previous investigations on this compound \cite{Minguet02, Lhotel06}. In Sec. \ref{results}, we present our magnetic measurements at very low temperature (below 800 mK), focusing on the avalanches detected in the hysteresis loops. Since these avalanches appear to be intrinsic to the system, we propose in Sec. \ref{discussion} that the observed behavior is related to  domain-wall propagation. We present a simple model that allows us to describe the main experimental features.

\section{Samples and magnetic characteristics}
\label{exp}
The studied compound is the  Mn(hfac)$_2$({\it R})-3MLNN (formula 
C$_{27}$H$_{25}$F$_{12}$MnN$_2$O$_9$) \cite{Minguet02} comprised of chains of Mn and nitronyl-nitroxide free radicals (NITR), called here Mn-NITR. We obtained our main results for three samples of different shapes and weights (see Table \ref{tabsample}). Samples 1 and 2 are needle shaped and were measured using low temperature SQUID magnetometers equipped with a miniature dilution refrigerator. The set-up can measure absolute values of the magnetization by the  extraction method \cite{Paulsen01}. Sample 3, about a thousand times smaller, was measured with a micro-SQUID magnetometer \cite{Wernsdorfer01}. All the measurements we present here were performed along the easy axis, the $b$ axis. In this direction, the values of the demagnetizing factor $N$ are quoted in Table \ref{tabsample}. They were deduced from magnetization measurements, sample shape and, in case of samples 1 and 2, ac susceptibility. 

\begin{table}[h]
\begin{tabular}{*{2}{c}r@{$\times$}c@{$\times$}lc}  \hline \hline
Sample&Mass&\multicolumn{3}{c}{Dimensions (mm)}&$N$ (cgs)\\ \hline 
Sample 1&1.68 mg&4.2&0.95&0.95&$\sim $1 \\ 
Sample 2&0.818 mg&4.7&0.6&0.5 & $\sim $0.6\\ 
Sample 3&$<$1 $\mu$g&0.1&0.05&0.05&$\sim$2.5 \\  
\hline \hline
\end{tabular}
\caption{\label{tabsample} Summary of measured samples. }
\end{table}

The magnetic structure of the compound is described in a separate paper \cite{Lhotel06}. It originates directly from the crystallographic structure \cite{Minguet02}. The zig-zag chains extend along the $b$-axis and are composed of alternating chiral NITR free radicals (({\it R})-3MLNN=({\it R})-Methyl[3-(4,4,5,5-tetramethyl-4,5-dihydro-1{\it H}-imidazolyl-1-oxy-3-oxide) phenoxy]-2-propionate)) carrying a spin $s_{\rm NITR}={1/2}$ and Mn(hfac)$_2$ units where the Mn(II) ion carries a spin $S_{\rm Mn}={5/2}$. The spins preferentially point along the chain axis.
This point, which is unusual in Mn(II) and radical based compounds, was adressed in Ref. \onlinecite{Lhotel06}where it was shown that the anisotropy cannot be explained by dipolar interactions: Single-ion anisotropy or antisymmetric exchange has to be invoked to account for the anisotropy \cite{footnote3}. The interaction between the Mn and radical spins along the chains is antiferromagnetic, very strong, $J_{intra} \sim 500$ K, and much larger than the interchain coupling $J_{inter}$ ($J_{intra}/J_{inter} \sim 10^4$). The magnetic susceptibility follows a purely 1D ferrimagnetic-like behavior at high temperature, crossing over to a 1D ferromagnetic behavior below 90 K involving effective spins $S_{tot}=S_{\rm Mn}-s_{\rm NITR}=2$.  An analysis of the susceptibility as a function of temperature in the range of 20 K$<T<$90 K allows us to estimate an equivalent ferromagnetic interaction $J=65$ K between the effective spins $S_{tot}$ \cite{Lhotel06}.

Below 20~K, interchain coupling effects become important, and give rise to a 3D long-range magnetic order below $T_c\sim $ 3 K. Magnetization and ac susceptibility measurements above $T_c$ suggested the existence of a ferromagnetic transition at  $T_c$. On the other hand, magnetization measurements performed below $T_c$ failed to show a spontaneous magnetization. Neutron diffraction measurements removed this contradiction. In zero field, the order is actually antiferromagnetic and is composed of alternating  planes of ferromagnetic aligned chains. However, the order is tenuous, and a  small field is sufficient to push the system into the ferromagnetic state (i.e., $\sim$ 150 Oe at 1.6 K) \cite{Lhotel06}. This behavior has been interpreted in terms of a subtle competition between the interchain correlations which are of the order of a few milliKelvins: long-range dipolar interactions which favor an antiferromagnetic order vs. weak short-range super-exchange interactions, responsible for ferromagnetic correlations (their presence was also inferred from measurements under pressure \cite{Laukhin04}): As the temperature is decreased, correlations along the chains grow, and the dipole field becomes proportionally stronger. At  $T_c$, it overwhelms the weak super-exchange interactions and 3D antiferromagnetic order is established. 

\section{Results: Dynamic properties}
\label{results}
\subsection{Hysteresis loops and magnetic avalanches}
\label{avalanches}
\begin{figure}[h]
\includegraphics[width=7cm, keepaspectratio=true]{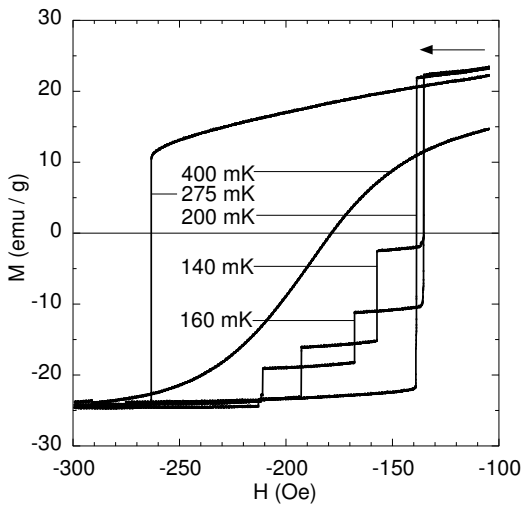}
\caption{$M$ vs. $H$: Decreasing part of the hysteresis loop for sample 1 at several temperatures between 140 and 400 mK at an equivalent field sweeping rate of 0.4 Oe.s$^{-1}$ (steps of 1 Oe every 2.5 s). The sample was saturated in a 2000 Oe field. Avalanches appear below 300 mK. }
\label{figMHaval}
\end{figure}

\begin{figure}[h]
\includegraphics[width=7cm, keepaspectratio=true]{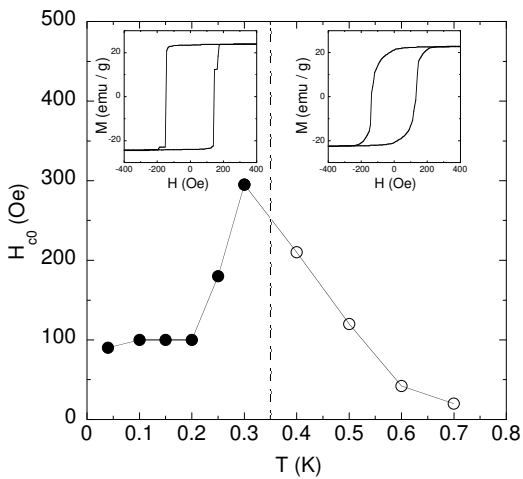}
\caption{Coercive field $H_{c0}$ as a function of temperature for sample 3 at a field sweeping rate of 350 Oe.s$^{-1}$. Empty circles: No avalanches occur. Full circles: $H_{c0}$ is equal to the first avalanche field. The line is a guide to the eyes. The insets show representative hysteresis loops in both regimes. }
\label{figHavalT}
\end{figure}

We focus here on the dynamic  properties in the 3D magnetically ordered phase below 800 mK, at temperatures where hysteresis loops occur. From measurements of the magnetization performed along the chain axis (which is  the easy magnetization direction), we perceive  two distinct  temperature regimes, readily apparent in the hysteresis behavior shown in Figs. \ref{figMHaval} and \ref{figHavalT}. In the high temperature regime (above 300 mK), the magnetization $M(H)$ decreases smoothly as the magnitude of the applied (negative) field increases. Below 300 mK, a new physics develops and one observes magnetic avalanches. In Fig. \ref{figMHaval}, we show the decreasing part of the hysteresis loops for a few illustrative temperatures for sample 1. The data are obtained after the application of a positive 2000 Oe field sufficient  to saturate the magnetization \cite{footnote1}. In the following, we define the coercive field $H_{co}$ as the field at which $M(H)$ crosses the  $M=0$ axis, whether avalanches are detected or not. The smooth hysteresis curves observed at high temperature give way to abrupt avalanches below 300 mK. Note that one avalanche is sufficient to totally reverse the magnetization at $T\geq$ 200 mK but several avalanches are needed for the same purpose below 200 mK (See Fig. \ref{figMHaval}). Nevertheless, in the latter case, the amplitude of the first avalanche was always  sufficiently large so as to cross the $M=0$ axis,  so  conveniently  $H_{co}$ is also the threshold field for occurrence of the first avalanche. The coercive field  depends on $T$, but not monotonously, in the studied temperature range. See the examples of sample 1 in Fig. \ref{figMHaval} and of sample 3 in Fig. \ref{figHavalT}. When decreasing the temperature, one observes that  $H_{co}$ first increases down to 300 mK (regime with no avalanches), subsequently decreases down to 200 mK and finally becomes temperature independent below 200 mK.  Here, we stress that the hysteresis features are really intrinsic, and sample independent. In fact, although the sample shapes and masses varied,  and different apparatuses and procedures were used for the measurements, we always found a maximum $H_{co}$ occurring at 300 mK. Also, below 300 mK, we always observed magnetic avalanches for all samples, regardless of the value of the field sweeping rate (See below). 

Once an avalanche takes place, it typically lasts for approximately 15~ms for the 1~mg Mn-NITR  samples. This leads to a 10-15 m.s$^{-1}$ propagation velocity, comparable to Mn$_{12}$ac avalanche velocity \cite{Paulsen95, Suzuki05_1}.  We could also detect a faint temperature pulse during the avalanche of a few millikelvins (the thermometer was approximately 10 cm away). This is, by comparison, much smaller (at least ten times) than what we previously observed in similar size Mn$_{12}$ac or CoTAC samples (taking into account the different avalanche field values).  We will see below that the characteristic internal field at which avalanches take place is $H_{1}$=170 Oe. Therefore, we can estimate the heat released for an effective spin flipping in this field: $\Delta T= g\mu_0 \mu_B \Delta S H_{1} / C_{LT} =1.9$ K (in which the low temperature specific heat $C_{LT} \approx 4\times 10^6\ \textrm{erg.mol}^{-1} \textrm{.K}^{-1}$ was measured by the quasi-adiabatic method down to 350 mK \cite{Lhotel_PhD}). Each effective spin is surrounded by six first neighbor spins located on adjacent chains. As a chain flips, we imagine  that the heat radiates away from the chain like spokes on a wheel. In first approximation, if all of this heat is absorbed by the nearest chains, it would result in a local heating that corresponds to a temperature rise of only  $\approx 300$ mK per spin. When compared to the characteristic energy scales of the system (see Sec. \ref{barrier}), this seems to indicate that local heating is not the only mechanism that contributes to the avalanche process. 

\begin{figure}[h]
\begin{center}
\includegraphics[width=7cm, keepaspectratio=true]{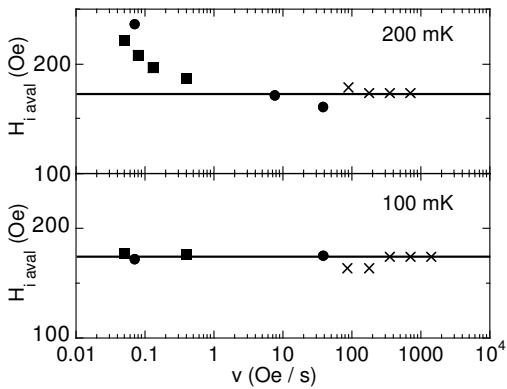}
\caption{First avalanche field $H_{i\ aval}$, corrected for demagnetizing field, as a function of field sweeping rate $v$, for two different temperatures, 100 and 200 mK. Squares, respectively circles and crosses correspond to sample 1, respectively sample 2 and 3. Lines indicate the field $H_1 =170$ Oe.}
\label{figHavalv}
\end{center}
\end{figure}

A crucial result concerns the value of the internal field $H_{i\ aval} $ for the first avalanche. $H_{i\ aval}$ is obtained from the applied field $H$ corrected for demagnetization effects in a mean field approach: $H_{i\ aval}=H-NM_b$, where $N$ is the demagnetization factor noted in Table \ref{tabsample} and $M_b$ is the value of the magnetization just before the occurrence of the avalanche. At 100 mK, and when the sample is in the saturated state, we found the same value of $H_1=H_{i\ aval}(M_b=M_{sat})= 170$ Oe for all samples whatever their shape or  volume is (varying by a factor $10^{3}$) and independent of the field sweeping rate $v$, that we varied by 4 orders of magnitude from 0.04 to 700 Oe.s$^{-1}$ (See bottom of Fig. \ref{figHavalv}). In addition, the value of  $H_1$ is very reproducible: When repeating hysteresis loops about 200 times for samples 1 and 2, we found the first avalanche field to be constant to within 2 \%. Furthermore, as seen in Fig. \ref{figMHaval}, below 200 mK, the reversal occurs in several avalanches. It is worth noting that, when corrected for demagnetization effects, the successive avalanches occur at the same internal field of 170 Oe (but with a larger distribution than the first avalanche field). These properties  lead us to conclude that this avalanche field is an intrinsic property of Mn-radical chains. 

At 200 mK, $H_{i\ aval}$ is still constant, but only for sweeping rates faster than 0.5 Oe.s$^{-1}$. For field sweeping rates slower than 0.5 Oe.s$^{-1}$,  $H_{i\ aval}$ is no longer constant (See top Fig. \ref{figHavalv}). In fact, the experimental conditions have changed when the field ramping is very slow at this higher temperature. This is because the magnetization has time to relax during the sweep due to thermally activated processes and therefore the magnetization $M_b$ just before the avalanche differs substantially from the saturation value. 
This suggests that the value of the magnetization just before the avalanche $M_b$  plays a  crucial  role in the determination of $H_{i\ aval}$. 

\begin{figure}[h]
\includegraphics[width=7cm, keepaspectratio=true]{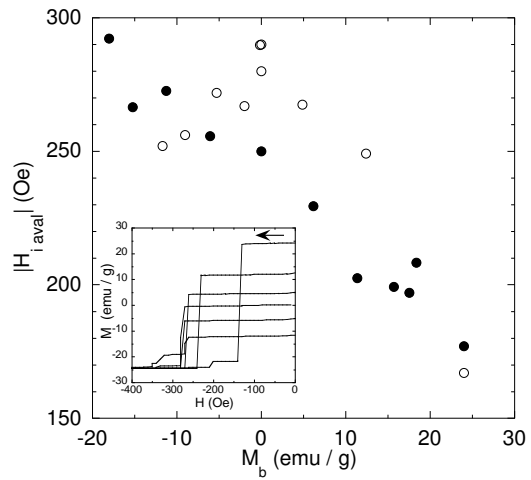}
\caption{The first avalanche field $H_{i\ aval}$ vs. the initial magnetization $M_b$ just before an avalanche for sample 2, at 100 mK and for field sweeping rates of 9.6 Oe.s$^{-1}$ (full circles) and 
0.12 Oe.s$^{-1}$ (empty circles). The inset shows M(H) curves obtained at 0.12 Oe.s$^{-1}$ for several $M_b$ values. }
\label{figHavalMb}
\end{figure}

To clarify this point, we performed a series of  M(H) measurements at 100 mK, starting from a non-saturated state. For these measurements, we field cooled the sample in various fields 
from above 2 K (where there is no hysteresis) down to 100 mK in order to freeze the sample in different given magnetization configurations, with different  amounts of initial magnetization. The resulting avalanche field $H_{i\ aval}$ at 100 mK as a function of the magnetization just before an avalanche $M_b$ is shown in Fig. \ref{figHavalMb} for two field sweeping rates.  As can be seen, $H_{i\ aval}$ increases with decreasing initial magnetization, and seems to saturate at about 300  Oe.  Note, in particular, that the avalanche field  starting from the zero field cooled (ZFC) magnetization $H_2=H_{\rm i\ aval}(${\small $M_b=0$}$) \approx 270 \pm 20$~Oe \cite{footnote2} is larger than the avalanche field of $H_1=170$ Oe obtained from the saturated state (See Fig. \ref{figHavalMb}). One can see why a constant avalanche field is observed at low temperature when starting from saturation:  As the temperature is decreased, relaxation during the hysteresis loop becomes increasingly slow. Below 300 mK, the starting value for $M_b$  remains very close to saturation, and according to Fig. \ref{figHavalMb}, $H_{i\ aval}$ will occur at approximately 170 Oe.

\subsection{Estimation of the characteristic energy barrier}
\label{barrier}

\begin{figure}[h]
\includegraphics[width=7cm, keepaspectratio=true]{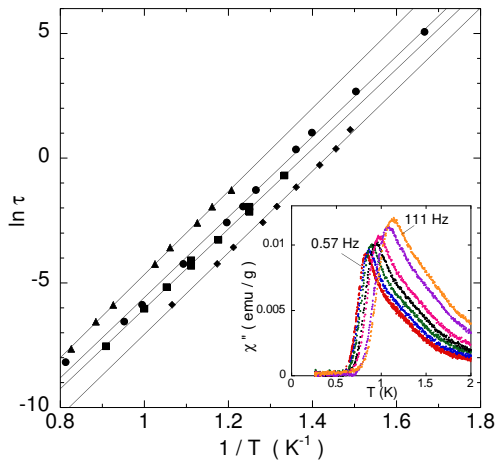}
\caption{(Color online) $\ln \tau$ vs. $1/T$ for four different samples, from ac susceptibility measurements with $H_{ac}=0.14$ Oe and 1.1 mHz$ < f < 511$ Hz ($\tau=1/2\pi f$). For each sample, the line represents the fit to an Arrhenius law: $\tau = \tau_0 \exp (E_0/k_BT)$, with the energy barrier $E_0/k_B \approx 16$ K and $7 \times 10^{-11}\ {\rm s} < \tau_0 < 5 \times 10^{-10}$ s. The inset shows the dissipative part $\chi"$ of the ac susceptibility as a function of temperature at frequencies between 0.57 and 111 Hz for a 1.59 mg single crystal.}
\label{figtau}
\end{figure}

The rapid variation of $H_{co}$ with temperature between 300 and 900 mK suggests an origin in terms of thermally activated processes. The latter can be well characterized from the study of the dependence of the magnetization on the measuring time $t$ (or frequency $f$) and temperature $T$. Thus, we studied the dissipative part $\chi"(T,f)$ of the ac susceptibility ($H_{ac}||b$) for samples cooled in zero static field (i.e. in the antiferromagnetic phase).  $\chi"(T,f)$ exhibits a maximum at a temperature $T_M$ that depends on the measuring frequency and obeys an Arrhenius law: $\tau=\tau_0 \exp(E_0 / k_BT_M)$, $\tau=1/2\pi f$ over six decades in frequency (See Fig. \ref{figtau} and Ref. \onlinecite{Minguet02}). The energy barrier $E_0$ is sample independent: For samples 1, 2 (but also for two other samples not presented here), we found the same value of $E_0/k_B= 16$ K even though the samples were synthesized in different batches over a two year period and had various thermal cycles. Only  $\tau_0$ is found to vary slightly between samples, ranging from $7 \times 10^{-11}$ to $5 \times 10^{-10}$ s , and may be explained by different characteristic chain lengths from sample to sample \cite{Bogani04}.  Cole-Cole plots of isothermal $\chi"(f)$ vs. $\chi'(f)$ are non-circular which implies a distribution of energy barriers $P(E)$ of which $E_0$ is the mean value. An analysis of our $\chi"(\omega)$ and $\chi'(\omega)$ data using models for the distribution of energy barriers \cite{Huser86} provides a rather narrow width of $P(E)$, of about $\Delta E_{1/2}/k_B=3\ K$ at half maximum  \cite{Lhotel_PhD}. 

\begin{figure}[h]
\includegraphics[width=7cm, keepaspectratio=true]{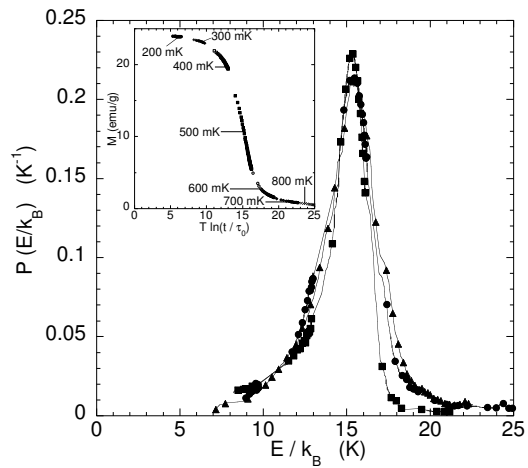}
\caption{Distribution of energy barriers, $P(E)$. Squares (circles): distribution obtained from relaxation vs. time $t$ at several temperatures on sample 1 (sample 2), Triangles: distribution obtained 
from relaxation vs. temperature $T$ of sample 2. The inset shows the magnetization $M$ as a function of $E/k_B=T \ln (t/\tau_0)$ with $\tau_0=2.8 \times 10^{-10}$ s from relaxation curves as a function 
of time of sample 2. All these relaxation curves were measured in zero field.}
\label{figP(E)}
\end{figure}

Below 800 mK,  $\chi"(T)$  vanishes. However, it is still possible to investigate $E_0$ and  $P(E)$ from 800 mK down to 300 mK by studying the relaxation of the saturated isothermal remanent magnetization $M_R$ in zero field. Note that during this relaxation process, the system passes from the field induced ferromagnetic phase to the  antiferromagnetic ground state. We did not find a simple exponential relaxation characteristic of the presence of a single barrier height. Instead, we found that the relaxation of $M_R$ depends on a single variable, $E_c (T,t)=k_BT \ln (t/\tau_0)$ and is characteristic of 
a distribution of barrier heights crossed by thermal activation. In this case, $P(E)$ can be found by adjusting the $M_R (T,t)$ data to  the formula $P(E=E_c)=dM_R/dE_c $ \cite{Prejean80}. Recording the evolution of the magnetization at different constant temperatures over 10 h following the removal of the field, we could superpose the $M_R(T,t)$ data on a single curve in the scaling diagram $M_R$ vs. $T \ln (t/\tau_0)$ when setting $\tau_0= 2.8 \times 10^{-10}$ s (See the inset of Fig. \ref{figP(E)}). From the fit, we deduce a $P(E)$ centered on the energy $E_0/k_B=$15.5 K with  $\Delta E_{1/2}/k_B=$3 K (See Fig. \ref{figP(E)}), thus in very good agreement to that deduced from our ac data in the antiferromagnetic phase above 800 mK. Thus, using different experimental methods (in different temperature ranges), we found nearly the same energy barrier  $E_0$ and distribution $P(E)$, independent of the sample used. 

Finally, we note that we also measured, after saturating the sample, the relaxation of the magnetization in a negative applied field.  Below 300 mK, and for field values less than $H_1$, we observed two regimes. At short times (beyond our experimental resolution), the magnetization shows a small decrease (few percent). At long times (up to 24 h), the relaxation is very slow, with a logarithmic tail, and at 100 mK less than a fraction of a percent of the magnetization relaxes.

\section{Discussion}
\label{discussion}
\subsection{Origin of the energy barrier}
In the following, we propose that the avalanche dynamics at low temperature are governed by the same energy barrier observed at higher temperature, which we argue arises from the energy required to nucleate a domain wall along a chain. However, the avalanche dynamics are found to be modified by the state of the magnetization $M_b$ just before the avalanche.  We suggest that it is a consequence of the competition between ferromagnetic and antiferromagnetic orders in Mn-NITR, and propose a scenario that allows us to describe the observed behavior by considering two situations: the avalanches from the saturated ferromagnetic state and those from the ZFC antiferromagnetic state. Finally, we discuss the questions that emerge from this scenario and note that the local heating and the subsequent diffusion of the heat front, although important, are not the driving forces behind the propagation of the avalanche.

We contend that the energy barrier $E_0/k_B \approx 16$ K  observed in measurements above 300 mK  is the energy required to nucleate a domain wall in the system. Although Mn-NITR orders three dimensionally and there are very strong 1D magnetic interactions within the chains, the  interactions between spins on neighboring chains (which ultimately result in the 3D order), are very weak, only of the order of a few millikelvins (See Sec. \ref{exp}). Because the interchain interactions  are so weak compared to the anisotropy along the chain, there can be no domain wall between chains. That is to say, for a given chain, the neighboring chains will  either be parallel  or anti-parallel, and thus for directions perpendicular to the chains, there is no "width" that one usually associates  with a domain wall.

On the other hand, as already evidenced in other spin chain compounds \cite{Lhotel06b,Balanda06,Evangelisti02}, along the chains the competition between the exchange and anisotropy energies results in the energetically favorable conditions for the nucleation of 0D domain-walls. To estimate the energy of such a domain-wall, we will, for simplicity, consider in this paper that the chains are made of effective spins $S_{tot}=2$. This is justified since all our experimental data below 20 K \cite{Minguet02, Lhotel06, Lhotel_PhD} are consistent with an effective spin approximation. Futhermore, as the NITR spin is delocalized, the spin distribution between the Mn and the NITR is not known \cite{Lhotel06}, making it very difficult for a more detailed description of the chain.Then, the cost in exchange energy at low temperature needed to flip an effective spin $S_{tot}=2$ on a chain edge would be J$S_{tot}^2$= 260 K, a very large energy. The cost to flip a spin inside a chain would be twice as much. However, the  nucleation of a 0D domain wall along a chain will have a much smaller activation energy. In a first approximation, we consider a 180$^{\circ}$ domain-wall with energy \cite{Kittel}: 

\begin{equation}
\label{Eparoi}
F=\frac{JS_{tot}^2 \pi^2}{2n}+\frac{nK}{2}
\end{equation}

where $n$ is the number of spins $S_{tot}$ in the domain-wall, $J$ is the exchange constant  between them, and $K$ is the anisotropy constant, supposing a uniaxial anisotropy.  Eq.~\ref{Eparoi} describes a domain-wall in which each spin is rotated by an angle of $\pi /n$ radians with respect to its neighbors. The first term is the cost in exchange energy for such a domain-wall, which decreases as $1/n$ and thus favors a more spread-out wall. The second term is the increase in anisotropy energy for such a domain-wall and is proportional to $n$. 

Magnetization isotherms and perpendicular susceptibility allow us to estimate the anisotropy  constant \cite{Lhotel06, Lhotel_PhD}: $K \approx 2 \times 10^4$ erg.cm$^{-3}$, that is to say 0.13 K$/$spin $S_{tot}$. The domain-wall energy $F$ (See Eq.~\ref{Eparoi}) is minimized for $n=\displaystyle \sqrt{\frac{JS_{tot}^2 \pi^2}{K}}=140$ effective spins in the chain. This rather large wall is a consequence of the weak anisotropy in this compound. The energy required to nucleate the wall is  $F/k_B=18$ K. This value is very closed to the experimental energy barrier $E_0/k_B =16$ K notwithstanding the simplicity of the model. Local defaults can modify this nucleation energy, which can explain the observed barrier distribution. Once a wall is nucleated, it will sweep along the chain reversing many more spins.

We can use this rough estimate of the number of spins in a domain wall in order to compare  the experimental  energy barrier  with the  Zeeman energy acquired by the wall just before avalanche takes place. Consider for example, the case when the sample has first been saturated in high field, and is thus in the ferromagnetic state. Avalanches always occur at $H_1=170$ Oe when $M_b=M_{sat}$. Assuming that the  same number of spins ($n=140$) is involved in the nucleation of the domain-wall, the Zeeman energy supplied by the field to these spins is $E_{H_1}/k_B=g \mu_B n S_{tot} H_1=6.4$ K (with $g=2$ from EPR measurements \cite{Vidal04}). In a similar fashion,  when the sample has been zero field cooled  in the antiferromagnetic state with $M=0$, avalanches take place when $H_2 \approx 270$ Oe. 
In this case, the Zeeman energy acting to flip the spins and create the wall is approximately 10.1 K. The above estimates of the energy scales, although systematically less than the experimental values, are nevertheless of the same order of magnitude.

Because the ferromagnetic state in zero field has a higher energy than the antiferromagnetic ground state in Mn-NITR, we might expect that it is easier to nucleate a domain wall in the ferromagnetic state. This may explain in part the dependence of the  Zeeman energy on the initial magnetization. Indeed, it has  been shown\cite{Lhotel06} that dipole interactions are responsible for the 3D antiferromagnetic phase transition at 3 K. So it seems reasonable to estimate the difference between the two states from magneto-static energy considerations. At $H=0$, the magnetic energy density  in the ferromagnetic state is -1/2 $ M_{sat} H_D =  740 $ erg.cm$^{-3}$ (where $H_D$ is the demagnetizing field with $N=1$). This corresponds to an increase of approximately 4.5 mK$/$spin 2 above the ground state. For a domain wall of 140 spins, this implies a difference in energy of about  0.65 K for the wall, significantly smaller than the observed difference of 3.7 K in the Zeeman energy for the walls. Intriguingly, however, this is of the same order of magnitude as the difference between the energy barriers measured in the antiferromagnetic state by the ac susceptibility  and the slightly smaller barrier  deduced from relaxation out of the antiferromagnetic state. 

\subsection{Phenomenological model}
\begin{figure}[h]
\includegraphics[width=8.5cm]{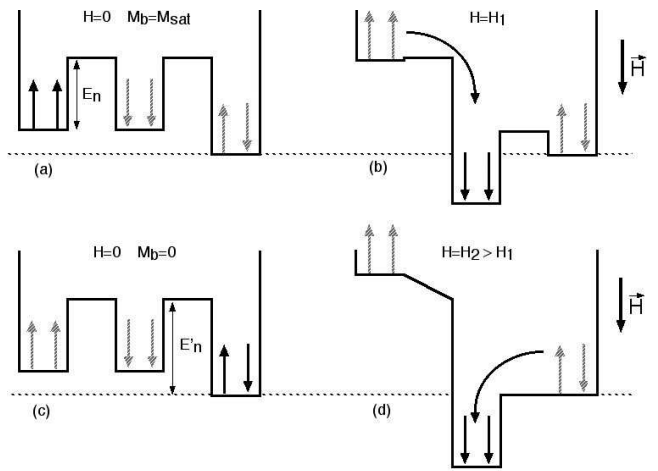}
\caption{Schematic diagrams of the energy barrier landscape. a) $M_b=M_{sat}$ and $H=0$: The system has been field cooled in the ferromagnetic-up state.  The nucleation energy barrier $E_n$ has to be crossed to go from the ferromagnetic-up state to the antiferromagnetic ground state. b) $M_b=M_{sat}$ and $H=H_1$: Because of the Zeeman energy, the ferromagnetic-down state has the lowest energy,  and $H_1$ is enough to suppress the barrier between the ferromagnetic-up and down states. The system avalanches into the ferromagnetic-down state. c) $M_b=0$ and $H=0$. The system has been zero-field cooled in the antiferromagnetic state which is separated from the others by the energy barrier $E'_n$. d) $M_b=0$ and $H=H_2>H_1$: The field is enough to suppress $E'_n$ between the antiferromagnetic state and the ferromagnetic-down state. The system avalanches into the ferromagnetic-down state.}
\label{schema}
\end{figure}

We suggest a simple model to describe qualitatively the observed behavior with the aid of  Fig. \ref{schema} which schematically shows the energy barrier landscape for two different initial states. In  Fig. \ref{schema}a, the sample has been first saturated in a high field and is in the ferromagnetic-up state at $H=0$. The ferromagnetic-up state and  ferromagnetic-down state are degenerate at $H=0$, but separated by an energy barrier. The antiferromagnetic ground state lies slightly below these. At  low temperature, thermal activation is not enough to overcome the energy barrier and the system remains in the excited ferromagnetic state. Fig. \ref{schema}b shows the effects of applying a magnetic field. In a negative field, the Zeeman energy of the ferromagnetic states is shifted with respect to each other: The up state increases in energy, the down state decreases by an equal amount, while the antiferromagnetic state remains the same. If the temperature is very low during the field ramp, relaxation is very slow and $M_b$ remains constant and close to the saturation value $M_{sat}$.  At the critical field $H_1$, the energy barrier of the up state will be reduced to near zero, and multiple nucleations of domain walls will occur, creating an avalanche. The system will slide down into the lowest state, which, due, to the applied field, is now the ferromagnetic down state. 

The situation when the sample has been zero field cooled in the antiferromagnetic $M_b=0$ ground state is shown in Fig. \ref{schema}c. In this case, a larger field must be applied in order to reduce the energy barrier and induce nucleations of domain walls, as shown in Fig. \ref{schema}d. The system then avalanches into the ferromagnetic-down state. Note that, in this simple model, it has been assumed that nucleation of domain walls occur when the energy barrier is reduced to nearly zero. Another possibility could be that the energy barrier remains finite, and  nucleation of the domain-walls takes place by quantum tunneling through the barrier.  This would, in effect, short circuit the energy barrier and explain why our estimates for the Zeeman energy of the walls are lower than the experimental energy barrier.

It is  interesting to look at the multiple avalanches at low temperature in light of our simple model. As seen in Sec. \ref{avalanches}, when corrected for demagnetization effects, the successive avalanches are found to occur at nearly the same internal field $H_1 \approx 170$ Oe. This result  implies that the sample remains in the ferromagnetic state during the avalanche process,  but is broken up into a few large ferromagnetic domains of opposing  polarity. This is in agreement with the scenario shown in Fig. \ref{schema}b. When avalanching out of the $M_{sat}$ state, the sample goes from the ferromagnetic-up state to the ferromagnetic-down state directly; i.e., it does not pass through the antiferromagnetic state. This also implies that the internal field is critical to the advancement of the ferromagnetic domain front. During the avalanche, the internal field is reduced below its critical value as the net magnetization changes. If the avalanche stops, then at the interface of two oppositely polarized domains, the field will be a minimum. Then, the external field has to be increased until the critical internal field is reached again, so that a new avalanche is induced.

This simple model does not address the apparent collective phenomena of the avalanche. This aspect is quite different from those previously observed in Mn$_{12}$ac SMM or the CoTAC chain. As the latter is also 3D ordered \cite{CoTAC}, it is interesting to compare:  As mentioned in the Introduction,  avalanches could be suppressed in CoTAC samples by simply ramping the field at a slower rate. At very low temperature, the relaxation of the magnetization from $+M_{sat}$ to $-M_{sat}$
could be measured. The cause of the relaxation was shown to be due to quantum nucleation of domain walls (made up of only ten spins or so) that took place at a resonant field of approximately 1000 Oe. Each nucleation of a domain wall by tunneling through an energy barrier could be treated as an independent event. 

This is very different from the present situation for Mn-NITR: At low temperature, avalanches always took place regardless of the sample size or slow ramping speed. At 100 mK, the relaxation of the magnetization was always extremely slow, $\tau > 10^6$ s, right up to $H_{i\ aval}$. Of course, when an avalanche took place, it was rapid,  $\tau \approx 10^{-2}$ s. After an avalanche, there was no further relaxation, which excludes the hypothesis of a resonance effect as  observed in CoTAC. Thus, there is an all or nothing aspect to the dynamics at low temperature of Mn-NITR.  Other important differences between the two systems includethe following: the spins in the Mn-NITR chains come from two different sources (the Mn(II) ion and from the delocalized electron of the free radical NITR) that are aligned along the chain direction, the 1D character is more than an order of magnitude stronger than in CoTAC, and the anisotropy is more than an order of magnitude weaker. 

These attributes  result in an  estimated domain wall length that is more than an order of magnitude longer in Mn-NITR in comparison with CoTAC. This leads us to speculate that  the importance of the long-range dipolar interchain interactions in Mn-NITR may provide the conditions for the collective nucleation of domain walls. Within the long and spread-out domain wall, a large number of spins will be perpendicular to the chain axis. This will  produce a sizable transverse field on the neighboring chains, which, in turn, should favor the inducement of a domain wall in the neighbors. 
 
That local heating operates in this system but does not dominate the avalanche process can be seen in the following illustration. Below 200 mK, avalanches always occurred in a few large, distinct steps, as can be seen in Fig. \ref{figMHaval} for $T=140$ mK.  The magnetic energy released when large blocks of spins flip, $2M H_{i\ aval}$ is converted to heat during the avalanche. However, at low temperature, this heat is not enough to raise the temperature high enough for thermal activation to be efficient and, thus, not enough to sustain the avalanche and  finish the sample off. On the other hand, above 200 mK, avalanches were always complete, that is to say, from $+M_{sat}$ to $-M_{sat}$ in one step. Thus, the same magnetic energy of $2M H_{i\ aval}$ generated during the avalanche, along with the additional thermal energy due to the higher starting temperature, was enough to overwhelm the sample and cause a complete magnetization reversal. 

\section{Conclusion}
In conclusion, we have shown that intrinsic avalanches occur in  Mn-NITR, which is unusual in macroscopic samples. Below 200 mK, they are characterized by a constant reversal field in the hysteresis loops, which does not depend on the sample size or batch, the temperature and the field sweeping rate. We propose that this avalanche field is related to the energy barrier measured in ac susceptibility and relaxation measurements and corresponds to the energy needed to nucleate domain-walls along the chains. We suggest that, by applying a field, we tilt the energy field landscape to the point that the barrier height goes to zero, and due to the purity of the sample and the sharp energy distribution, many domain-wall nucleations occur, resulting in an avalanche. We also speculate that this is a collective phenomena.

\acknowledgments We thank M.-A. M\'easson and D. Braithwaite for the specific heat measurements and W. Wernsdorfer for the micro-SQUID measurements. We also acknowledge J.-J. Pr\'ejean for continuous and fruitful discussions. This work has been supported in part by the European Commission under the Network of Excellence MAGMANet (contract 515767-2), by the Ministerio de Educaci\'on y Ciencia (Spain), under the project ejeC-Consolider CTQ2006-06333/BQU, and by Generalitat de Catalunya (2005SGR00362). Support from the R\'egion Rh\^one-Alpes  through the "Programme de Recherche Th\'ematiques Prioritaires" is also  gratefully acknowledged.


\begin{thebibliography}{99}
\bibitem{footnote3} The observation of anisotropy in Mn(II) and radical based compounds, although unusual, has been previously reported: see, for example, references 11,37,38 and 39 of Ref. \onlinecite{Lhotel06}.
\bibitem{footnote2} The avalanche field distribution is much larger when starting from the ZFC state. This may be due to the presence of domains which induces a strong inhomegeneity in the magnetization and so a large distribution of the internal field inside the sample.
\end{thebibliography}
\end{document}